\documentclass[superscriptaddress,showpacs,twocolumn,aps,pre,10pt]{revtex4-1}
\usepackage{graphicx,dcolumn,bm,bbm,amsmath,amssymb,color}

\newcommand{\eeq}{ \end{equation} }
\newcommand{\beq}{ \begin{equation} }
\newcommand{\bea}{\begin{eqnarray}}
\newcommand{\eea}{\end{eqnarray}}
\newcommand{\ga}{\alpha}
\newcommand{\gb}{\beta}

\newcommand{\bhu}{ \hat{\bf u} }

\newcommand{\bbr}{ {\bf r} }

\newcommand{\kbt}{k_{\rm B}T}

\begin{document}
\title{Unusual swelling of a polymer in a bacterial bath}

\author{A. Kaiser}
\email{kaiser@thphy.uni-duesseldorf.de}
\affiliation{Institut f\"ur Theoretische Physik II: Weiche Materie,
Heinrich-Heine-Universit\"at D\"{u}sseldorf,
Universit{\"a}tsstra{\ss}e 1, D-40225 D\"{u}sseldorf, Germany}
\author{H. L\"{o}wen}
\affiliation{Institut f\"ur Theoretische Physik II: Weiche Materie,
Heinrich-Heine-Universit\"at D\"{u}sseldorf,
Universit{\"a}tsstra{\ss}e 1, D-40225 D\"{u}sseldorf,
Germany}

\date{\today}


\begin{abstract}

The equilibrium structure and dynamics of a single polymer chain in a thermal solvent
is by now well-understood in terms of scaling laws. Here we consider a polymer
in a bacterial bath, i.e. in a solvent consisting of active particles which bring in nonequilibrium
fluctuations.  Using computer simulations of a self-avoiding polymer chain in two dimensions
which is exposed to a dilute bath of active particles, we show that the Flory-scaling exponent
is unaffected by the bath activity provided the chain is very long.
Conversely, for shorter chains, there is a nontrivial coupling between the bacteria intruding into the chain
which may stiffen and expand the chain in a nonuniversal way. As a function of the molecular weight, the swelling
first scales faster than described by the Flory exponent, then an unusual plateau-like behaviour is reached
and finally a crossover to the universal
Flory behaviour is observed. As a function of bacterial activity, the chain end-to-end distance
exhibits a pronounced non-monotonicity.
Moreover, the mean-square displacement of the center of mass of the chain shows a ballistic behaviour 
at intermediate times as induced by the active solvent.
Our predictions are verifiable in two-dimensional bacterial
suspensions and for colloidal model chains exposed to artificial colloidal microswimmers.
\end{abstract}

\maketitle

\section{Introduction}
\label{sec:intro}

The physics of polymer chains in  a thermalized bath is governed by scaling laws.
One of the most fundamental scaling relates the typical extension of a polymer chain $R$
to its molecular weight $N$ culminating in the traditional Flory exponent $\nu$, such that
$R\propto N^\nu$ \cite{deGennesbook,Doi_Edwards_book} for very large $N$.
While $\nu=1/2$ for a Gaussian chain, a self-avoiding chain exhibits a Flory exponent $\nu > 1/2$ 
which depends on the spatial dimensions
$d$, we have $\nu=0.588 \approx 3/5$ in three and $\nu=3/4$ in two
dimensions \cite{book_by_Lothar_Schaefer}. Similar scaling laws apply to the polymer dynamics
where hydrodynamic interactions between the monomer play a crucial role \cite{Doi_Edwards_book}.
It is important to note that these basic considerations are designed for equilibrium situations,
i.e.\ the solvent is a thermal bath at temperature $T$ and the chain is not exposed to external fields.

In this paper we consider a polymer chain in a bacterial (or active) bath which consists of swimming
particles or bacteria. The collisions of the bacteria with the chain lead to nonequilibrium
(non-thermal) fluctuations of the chain which may result in new phenomena of chain stretching and compaction
different from equilibrium solvents. Active matter itself has been intensely explored over the last years,
both for living systems as bacteria \cite{2007SoEtAl}, spermatozoa \cite{2005Riedel_Science} and mammals
\cite{Vicsek_Report2012,Schadschneider} or is system of
artificial microswimmers \cite{Paxton,snezhko_nature,BtH_L_part,BocquetPRL2010,Bibette,MagnetoSperm} 
with various propulsion mechanisms \cite{Dietrich,Kapral2010,Bechinger_SM11,PalacciJACS} and a plethora of
nonequilibrium pattern formation phenomena were discovered
\cite{Narayan,2012SwinEtAl,Menzel_EPL,Clement_NJP14,2011Herminghaus,Golestanian14CometSwarm,PNAS,2008Saint_Shelley,ISA-PNAS14,Li_PRE,Palacci_science,Bialke_PRL2013,Baskaran_PRL2013,MarchettiPRL12,ZoettelPRL14,Egorov}.
At fixed system boundaries active system show distinct clustering and trapping behaviour
\cite{Wensink2008,Reichhardt_PRL2008,MarconiPRE,ElgetiGompper13,Lee13Wall,HaganConfinement,Chaikin2007,TrappingSperms,Kaiser_PRL,CatesTrapping,TrappingPeruani,Trapping_OGSchmidt}
and can be expoited to steer the motion of microrotors and microcarriers \cite{KaiserSokolov_2014,SokolovPNAS,LeonardoPNAS}
of fixed shape.

Here we link the field of microswimmers to polymer physics and consider a single
polymer chain in a bacterial bath (or an active solvent). The motivation to do so is threefold:
first, from a fundamental point of view, there is a need to understand how polymer scaling laws are
affected by non-bulk or nonequilibrium situations \cite{Netz2012,WnklerPolyShearFlow06,Muthukumar2004,WinklerGompper2008}.
An active solvent which is intrinsically
in  nonequilibrium is one of basic cases which put the scaling laws into questions. Second,
the collective behaviour of microswimmers has been studied at moving boundaries 
\cite{KaiserPopowa_2013,KaiserSokolov_2014,AngelaniCARGO} but all of
which were of fixed shape. Bacteria and active particles near flexible boundaries have not yet been explored
systematically and it is interesting to understand how clustering and trapping phenomena are modified
for flexible boundaries \cite{ElasticConf}. Our case of a flexible polymer chain is therefore one of the simplest key examples
to proceed along this important direction.
Third, in general, the set-up we are proposing is realizable in experiments and  relevant
for biological systems where swarms of bacteria are moving close to flexible objects
like at water-air interfaces \cite{Marschall,Pitt,VanDerMei,Mills,Berenike_Maier,Ariel}.
Our two-dimensional model can indeed be realized e.g. by inserting long polymers into two-dimensional
Bacillus subtilis suspensions \cite{2012SwinEtAl,PNAS,KaiserSokolov_2014}. Another complementary realization is by
exposing colloidal model chains \cite{Pine_Nature_2010}
to artificial colloidal microswimmers \cite{SenPNAS13,BocquetPRL2010,Bechinger}.

We use computer simulations of a self-avoiding polymer chain in two dimensions
which is exposed to a dilute bath of active particles. As a result, we show that the  Flory-scaling exponent
$\nu =3/4$ is unaffected by the bath activity provided the chain is very long.
For shorter chains, there is a nontrivial coupling between the bacteria intruding into the chain
which  stiffen and expand the chain. As a function of the molecular weight, the swelling
first scales faster than described by the Flory exponent until a plateau-like behaviour with a slight 
non-monotonicity is reached. This is nonuniversal
behaviour which reminds to the swelling of polymers in quenched disordered where similar
nonmonotonicities have been observed \cite{ChandlerJCP92,ChandlerJCP95} which have, however, a different physical origin.
Finally, for large molecular weights,  a crossover to the universal
Flory behaviour is observed. Moreover, as a function of bacterial activity, the chain end-to-end distance
shows a pronounced non-monotonicity.
The dynamical correlations exhibit a diffusive behaviour for very short and long times
in qualitative accordance with an equilibrated polymer, while an intermediate ballistic regime
can be found in the mean-square displacement of the center of mass of the chain induced by the active solvent.

This paper is organized as follows: we introduce our model and our computer simulation technique
in section \ref{sec:model}. Various results on the statistics of polymer structure in a bacterial bath are presented
in section \ref{sec:results} while the polymer dynamics is discussed in section \ref{sec:DynScaling}. Finally we 
conclude and give an outlook in section \ref{sec:conc}.

\section{Model}
\label{sec:model}

\begin{figure}[thb]
\begin{center}
\includegraphics[width=1\columnwidth]{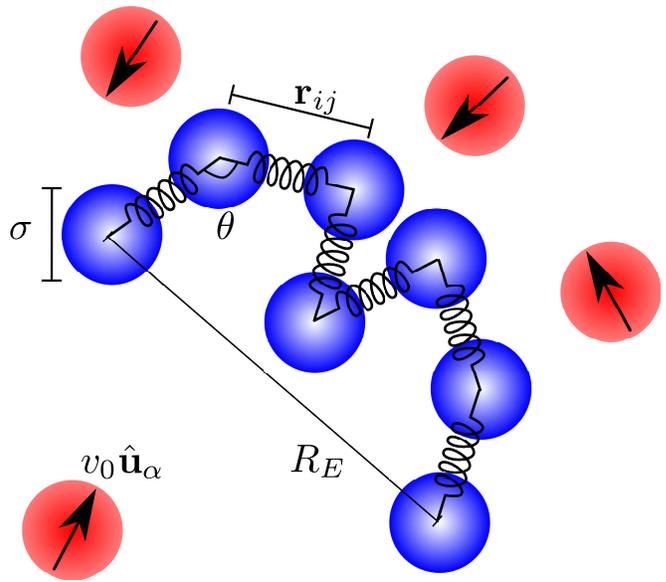}
\caption{\label{Sketch} Schematic sketch of the system for a chain with a low number $N=7$ of beads (blue) 
connected by springs with bond angle $\theta$. Furthermore the end-to-end distance $R_E$  is indicated. 
The $\alpha$th self-propelled disk (red -- bottom left) is driven along the marked orientation 
$\bhu_{\ga}$ (black arrow) with an velocity $v_0$.}
\end{center}
\end{figure}

We study the statistics of a polymer chain, modeled as a sequence of $N$ coarse-grained spring beads, in a bacterial bath,
composed of spherical swimmers in two dimensions, see Fig. \ref{Sketch}.
For simplicity, interactions between the active particles and the chain as well as inter-chain interactions are modeled by the same repulsive WCA-potential

\bea
U_{\text{WCA}}(r) = 4 \epsilon \left[ \left(\frac{\sigma}{r}  \right)^{12} - \left( \frac{\sigma}{r}  \right)^{6}  \right] + \epsilon,
\eea

for distances $r < 2^{1/6}\sigma$. Here the diameter of a bead and a disk-like swimmer is assumed to be equal 
and is denoted with $\sigma$ and $\epsilon=\kbt$ is the interaction strength. These quantities represent the length and
energy units, while times are measured
in $\tau = \sigma^{2}/D_0$, where $D_0$ is the short-time diffusion constant of a single monomer.

Springs are introduced via a socalled FENE (finitely extensible nonlinear elastic) potential \cite{KremerMDPolymer1990}

\bea
U_{\text{FENE}}(r_{ij}) = - \frac{1}{2}K R_{0}^{2} \ln \left[ 1- \left(\frac{r_{ij}}{R_0} \right)^{2} \right],
\eea

with neighboring beads $i,j$ and their distance $r_{ij} = |\bbr_i - \bbr_j |$. The spring constant is
fixed to $K=27\epsilon/\sigma^{2}$ and the maximum allowed bond-length to $R_0 = 1.5\sigma$. These interactions 
ensure that for the parameters chosen the swimmers do not cross the polymer chain.

In our chosen units, the overdamped equation of motion of the $i$th bead located at position $\bbr_i = [x_i(t),y_i(t)]$ is given by

\bea
\partial_t \bbr_{i}(t) &=& - \nabla_{i} U + \boldsymbol{\xi}_{i}
\eea

where $\boldsymbol{\xi}_{i}$ is Gaussian white noise with zero mean and correlations
$ \langle \boldsymbol{\xi}_{i} (t) \boldsymbol{\xi}_{j}(t^{\prime}) \rangle = 2 D_0 \delta_{ij}
\delta(t - t^{\prime}) \mathbbm{1}$ with the unit tensor $\mathbbm{1}$, and $U$ is the total potential energy.
The overdamped equation of motion for a swimmer $\ga$ is described through

\bea
\partial_t \bbr_{\alpha}(t) &=& - \nabla_{\ga} U +  v_{0} \bhu_{\alpha}(t) + \boldsymbol{\xi}_{\ga}
\eea

where $\boldsymbol{\xi}_{\ga}$ is Gaussian white noise as before, $U$ is the total potential interaction,
and $v_0$ is a self-propulsion velocity directed along
$\bhu_{\alpha} = \left(\cos \varphi_\ga, \sin \varphi_\ga \right)$, which will be given by the
dimensionless P\'{e}clet number $Pe = v_0 \sigma / D_0$. The presence of $v_0$ brings the system inherently into non-equlibrium.
Finally, the orientation of the swimmer is coupled to the rotational Langevin equation

\bea
\partial_t \bhu_{\alpha}(t) &=& \boldsymbol{\zeta}_{\ga} \times \bhu_{\alpha}(t).
\eea

Here $\boldsymbol{\zeta}_{\ga}$ is as well a Gaussian-distributed noise with zero mean and variance
$ \langle \boldsymbol{\zeta}_{\ga} (t) \boldsymbol{\zeta}_{\gb}(t^{\prime}) \rangle = 2 D_r \delta_{\ga \gb}
\delta(t - t^{\prime}) \mathbbm{1}$ and the corresponding rotational diffusion coefficient is $D_r = 3D_0 / \sigma$.

Steric interactions between the active particles are modeled by a soft repulsive Yukawa potential.
The total pair potential between a pair of disks $\{\ga,\gb\}$, is given by

\bea
U_{\ga \gb} = U_{0} \frac{\exp(-r_{\ga \gb}/\sigma)}{r_{\ga \gb}},
\eea

where the screening length corresponds to the disk diameter $\sigma$ and $r_{\ga \gb} = | \bf{r}_{\ga} - \bf{r}_{\gb} |$ 
is the distance between the swimmers, the prefactor is set to $U_0 = 20\epsilon$.

We perform Brownian dynamic simulations for various chain lengths $1 \leq N \leq 1000$
($N=1$ refers to the case of a single spherical tracer) using periodic
boundary conditions in a square simulation domain with an area $A=L_0^2$ where $L_0 \sim N \sigma$ corresponds
to the contour length of a linear chain in equilibrium.
In integrating the Brownian dynamic equations of motion, we have used a finite time step $10^{-7} \tau$.
The number of bacteria is determined by the dimensionless packing fraction
\bea
\phi=\frac{N_S \sigma^2}{4A},
\eea
where $N_S$ is the number of swimmers. We are interested in a dilute bacterial bath, so we chose $\phi \leq 0.02$,
which is below the jamming transition for self-propelled disks \cite{WensinkJPCM}.
Statistics are gathered for 20 to 50 independent simulation runs.

\section{Results}
\label{sec:results}
\begin{figure}[thb]
\begin{center}
\includegraphics[width=1\columnwidth]{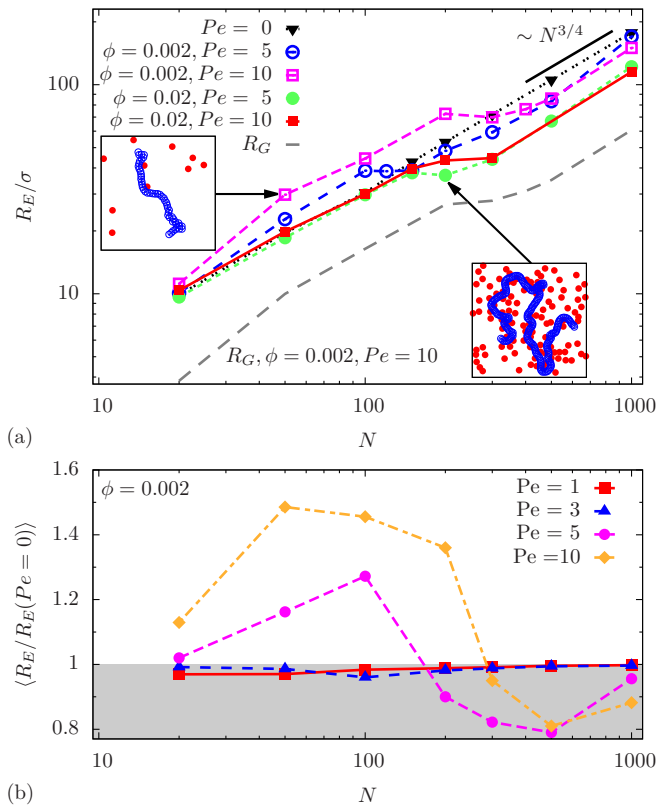}
\caption{\label{Flory} (a) End-to-end distance $R_E$ as a function of number of monomers $N$ for different bacterial
densities $\phi$ and different self-propulsion strengths $Pe$. Line corresponds to the
radius of gyration $R_G$ of the polymer for the given parameters. (b) Relative end-to-end distance, now scaled
with its value at vanishing P\'eclet number, versus molecular weight $N$ for different P\'eclet numbers at $\phi=0.002$.}
\end{center}
\end{figure}

As a key result, in Fig. \ref{Flory} the dependence of the end-to-end distance on the molecular weight is shown on a 
double-logarithmic plot where the slope indicating the typical two-dimensional Flory scaling with $\nu =3/4$ is also indicated. 
For small chain lengths, the polymer swells stronger than Flory scaling
which is obviously more pronounced for large activities $Pe$.
The strong swelling results from events where a bacterium intrudes into the polymer chain stretching it, 
see inset of Fig. \ref{Flory}. Increasing the
molecular weight $N$ further results again in more coiling such that a plateau-like regime is reached, the 
associated molecular weight needed to reach the plateau depends on the P\'eclet number.
Even a slight nonmonotonicity is compatible with the statistical uncertaincies. Finally a crossover to the universal 
Flory behaviour of a self-avoiding chain is observed. This is expected since at very large scales only the statistics 
of self-avoidance should matter. In this limit, the presence of the bacteria
are just providing some kind of higher effective temperature to the polymer such the typical entropically generated Flory
exponent is obtained, compare to a similar finding in \cite{Bialke_PRL2012}. Clearly, the non-universal plateau-like behaviour is
also found when the radius of gyration is plotted instead of the end-to-end distance, see again Fig. \ref{Flory}.
In Fig. \ref{Flory}(b), the polymer extension is again shown versus the molecular weight $N$ but is now scaled 
with its equilibrium value for vanishing activity
at same $N$. By definition, this quantity is unity when $Pe=0$ but varies with increasing P\'eclet number. 
Interestingly, for this quantity there is a marked nonmonotonicity in $N$ at intermediate P\'eclet numbers. 
For small $N$, the scaled end-to-end distance $R_E/R_E(Pe=0)$ is larger than unity quantifying the stretching 
effect by the bacteria sliding along the polymer chain. For larger $N$,
the collisions of the bacteria with the polymer chain lead effectively to a compression as signalled by $R_E/R_E(Pe=0)<1$.

\begin{figure}[thb]
\begin{center}
\includegraphics[width=1\columnwidth]{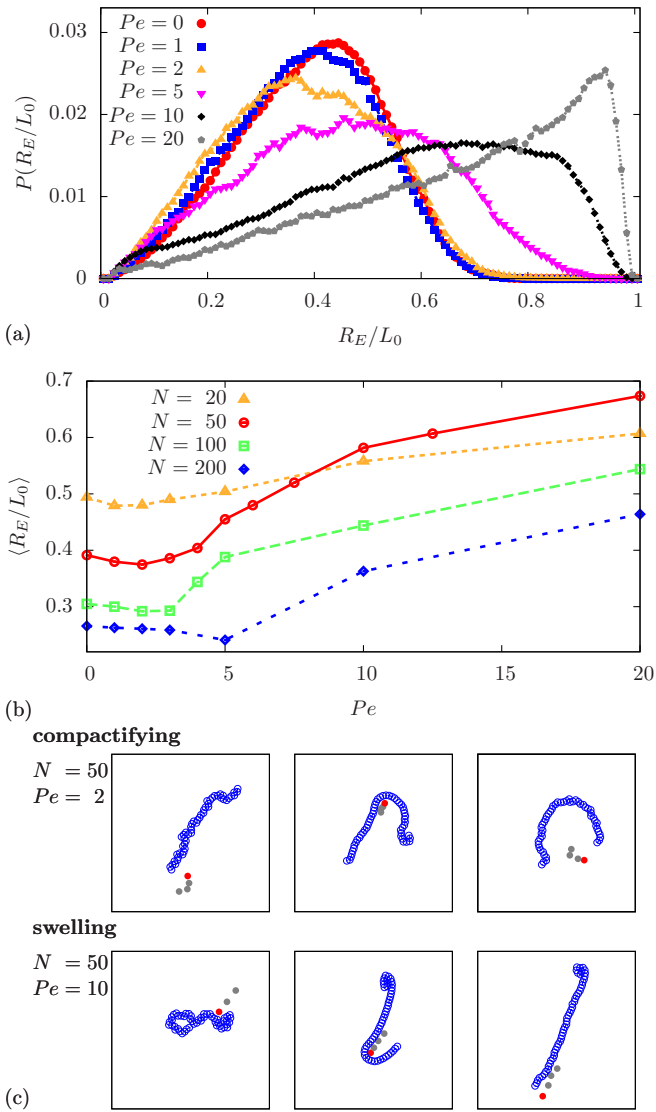}
\caption{\label{RE} (a) Probability distribution function of the reduced end-to-end distance $R_E/L_0$ for various
self-propulsion strengths $Pe$ and fixed chain length $N=50$. (b) Averaged reduced end-to-end distance $\langle R_E/L_0 \rangle$
as a function of $Pe$ for various chain lengths and end-to-end distance for various molecular weights $N$.
(c) Time sequences showing the compactifying and the swelling of a polymer due to an active swimmer.
Swimmer trajectories are indicated by its swimmer positions.}
\end{center}
\end{figure}

The distribution of the end-to-end distance is shown in Fig. \ref{RE}(a) for various P\'eclet numbers for a fixed molecular 
weight $N=50$ revealing a broad peak which  first shifts to the left and subsequently to the right for increasing $Pe$.
For intermediate P\'eclet numbers the peak is pretty broad documenting the strong polymer fluctuations imprinted by the 
bacterial bath. The quantitative analysis of the shift is shown in Fig. \ref{RE}(b) where
the end-to-end distance in units of the contour length $L_0$ is plotted versus P\'eclet number for various $N$ 
and at fixed diluted $\phi=0.002$.
For fixed $N$, a nonmonotonic behaviour of $R_E/L_0$ (or equivalently of $R_E/\sigma$) is clearly revealed 
as a function of P\'eclet numbers.
This can qualitatively understood as follows: for small $Pe$ a bacterium intrudes into the swollen chain 
and thus compactifying it, see also the snapshot series in Figure \ref{RE}(c). The larger Pe becomes, the more is the 
bacterium able to really stretch the chain by sliding along it which then induces an increase in the
averaged polymer extension. This scenario occurs over the whole range of molecular weight explored in this paper and is 
therefore quite general. The critical P\'eclet number for which the averaged polymer size is getting minimal increases 
with increasing molecular weight $N$, see Fig. \ref{RE}(b).

\begin{figure}[thb]
\begin{center}
\includegraphics[width=1\columnwidth]{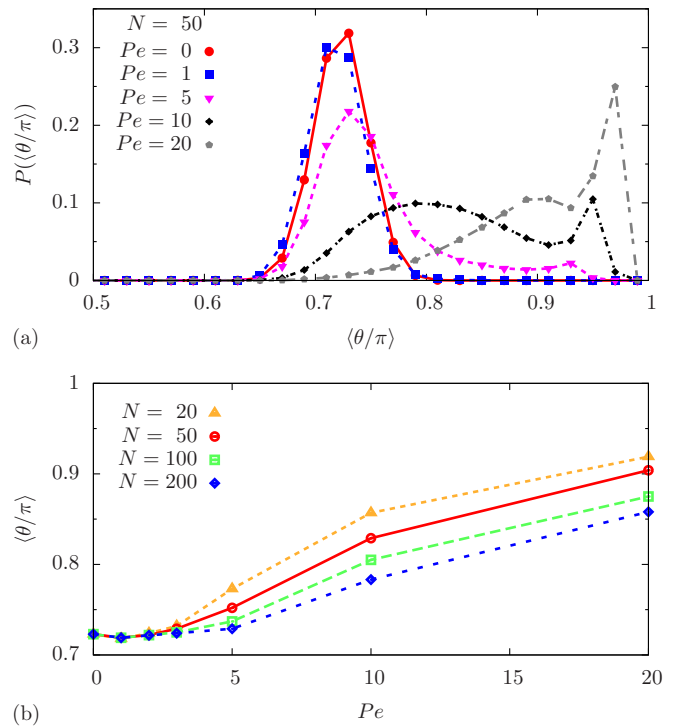}
\caption{\label{Ang} (a) Probability distribution function of the averaged bond angle 
$\langle\theta \rangle$ at fixed $N=50$ for various P\'eclet numbers $Pe$. 
(b) Averaged bond angle $\theta$ versus self-propulsion strength $Pe$ for 
various molecular weights $N$.}
\end{center}
\end{figure}

Finally we explore the impact of the intruding bacteria on the bond angle $\theta$ of subsequent monomers along the 
chain in Fig. \ref{Ang}. 
The statistical distribution of $\theta$, as shown in Fig. \ref{Ang}(a), reveals a double peak of stretched parts of the 
chain where the bacteria are scratching along and a coiled part unaffected by the bacteria, see again Fig. \ref{RE}(c).
 The average value $\langle \theta \rangle$ increases with $Pe$ reaching slowly the asymptotic value of 
$\pi$ due to full chain stretching induced by a bacterium travelling along the polymer, see Fig. \ref{Ang}(b).

\section{Polymer dynamics}
\label{sec:DynScaling}

We finally turn to the influence of the bacterial bath on the polymers dynamics which is typically measured in 
terms of mean-square displacements.
One may consider the latter for the end-monomer position $\bbr_N$,
the end-to-end distance $R_E$ itself, and the center of mass position $\bbr$.

Let us first recall the well-known scalings for a polymer in a thermal bath, corresponding to the case $Pe=0$.
In equilibrium, in the absence of hydrodynamic interactions,
the mean square displacement of the {\it end-monomer} behaves as \cite{NetzEndmonomer}

\bea
\langle \left( \Delta \bbr_N \right)^{2} \rangle \sim t + \lambda (1- \exp^{-t/\tau_p}),
\eea
for long times where $\lambda$ is a constant coefficient and $\tau_p$ a characteristic
 polymeric relaxation time. For very short times $\langle \left( \Delta \bbr_N \right)^{2} \rangle$
is diffusive (i.e. linear in $t$). The crossover behaviour from short to long times can be studied in terms
of the logarithmic derivative which sets an effective time-dependent exponent
as $\gamma(t) = d \log \left( \Delta \bbr_N \right)^{2} / d \log t$.
As a function of increasing time $t$, this exponent first decreases from 1 down to values of approximately $0.5$
and then increases back to 1. 

The mean-square displacement of the {\it end-to-end distance} in equilibrium is given for long times by \cite{Netz2012}
\bea
\langle \left( \Delta R_E \right)^{2} \rangle \sim (1- \exp^{-t/\tau_p})
\eea
such that it approaches it limiting values exponentially in time,
while it is again diffusive for short times and approximately scales with 
$\langle \left( \Delta R_E \right)^{2} \rangle \sim t^{1/2}$ for intermediate times \cite{Netz2012}.

Finally the mean square displacement of the {\it center of mass} of the polymer chain scales in equilibrium as

\bea
\langle \left( \Delta r \right)^{2} \rangle \sim t
\eea
which turns out to be a good approximation for all times.

\begin{figure}[thb]
\begin{center}
\includegraphics[width=0.9\columnwidth]{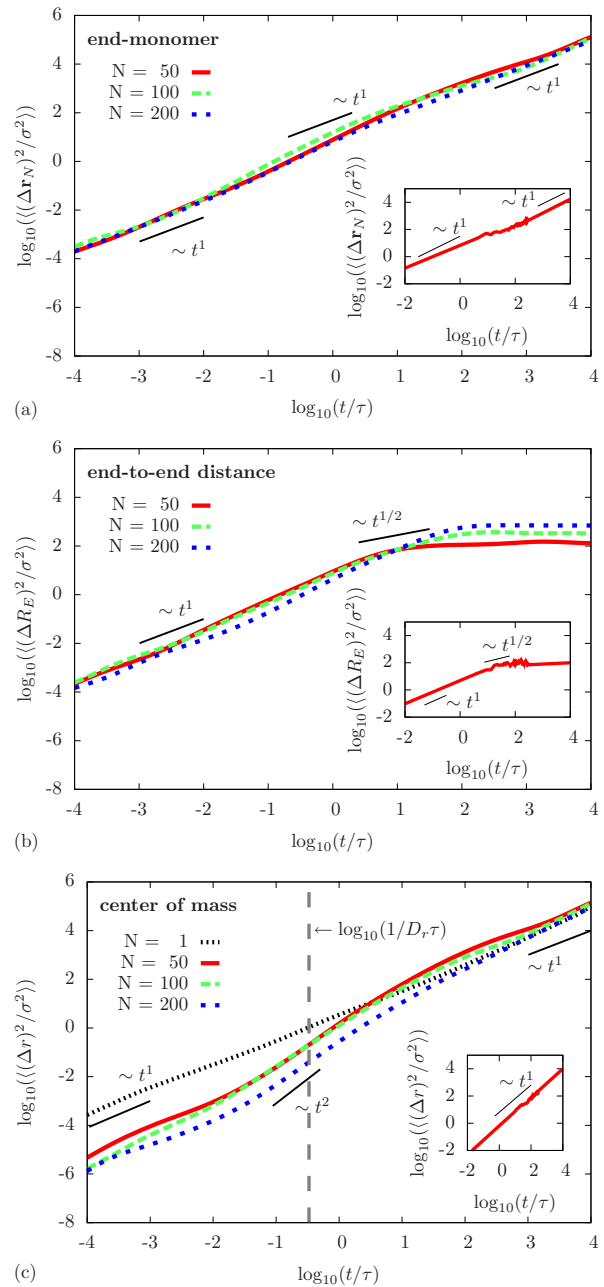}
\caption{\label{MSDs} Mean square displacements of (a) the {\it end-monomer}, (b) the {\it end-to-end distance},
and (c) the {\it center of mass} for various chain lengths $N$ and a swimmer density of $\phi=0.02$ with
self-propulsion strengths $Pe = 10$.
The insets show the temporal behavior for a polymer in a pure thermal bath. The dashed vertical line in (c) 
indicates the rotational diffusion time scale $1/D_r$ for an active swimmer.}
\end{center}
\end{figure}

In Fig. \ref{MSDs} we compare these well known mean-square displacements for a polymer chain in a thermal
bath (insets) with those for a chain in a bacterial bath with $\phi=0.02$ at a self-propulsion
strength $Pe=10$.
We observe for all three studied quantities the same short and long-time
behaviour. For short time, this is simply a result of our Brownian model. For long times, it is expected that
an active solvent can hardly be discriminated from a passive one on average.
At intermediate time, however, different behaviour gets visible. First of all,
the end-monomer mean-square displacement shows an acceleration at intermediate times
resulting in a larger value for the exponent $\gamma (t)$, compare Figure Fig. \ref{MSDs}(a) with its inset.
This obviously has to to do with the intruding bacteria which brings in more dynamics into the chain.
This effect is less pronounced for the
end-to-end dynamics (Fig. \ref{MSDs}(b)) where the dynamical behaviour is qualitative similar 
to the passive case, see the inset in Fig. \ref{MSDs}(b).

Conversely, for  the center of mass motion, there is a strong amplification of the bacterial dragging effect on the chain.
 Clearly, even new intermediate ballistic scaling regime shows up here
where the mean-square displacement scales as $t^2$, compare Fig. \ref{MSDs}(c) with its inset. 
Again, this has to do with the fact that in this regime
the active particles drags the whole chain with it. The ballistic regime  typically
ceases to exist when the particle decorrelates its orientation, i.e. its pulling or dragging force, which occurs on a time scale
$1/D_r$. This has been studied in great detail for a single Brownian active particle
\cite{BtH2011,Howse_2007,Kaiser_PRE2013Janus}.
The decorrelation time scale $1/D_r$  is plotted as a reference 
in Fig. \ref{MSDs}(c) and represents indeed a reasonable upper bound 
at which the ballistic regime ceases to exist.
Finally, as a further extreme reference, we have included the case $N=1$ of a single segment, representing a
 passive tracer in a bath of active particles as studied recently \cite{TracerGoldstein,TracerClement}. 
In this case, the ballistic regime is not very visible, since the collison time of active particles with the tracer is short.

\section{Conclusions}
\label{sec:conc}

We have considered to impact of an active (bacterial) bath on the conformations of a flexible polymer chain in two 
dimensions by using extensive Brownian dynamics computer simulations. While the traditional Flory scaling for a 
two-dimensional self-avoiding random walk is found for large chains, there is an interesting nonuniversal behaviour 
for finite chain lengths. Due to the intruding bacteria, the polymer extensions is getting
more stretched than predicted by Flory scaling and then crosses over to a plateau where the chain size does not 
depend on the molecular weight.
This behaviour is unusual as it is not found in equilibrium. We have further identified trends of the chain size with
increasing bacterial activity and find a relative compression for small activities and a strong stretching for 
large activities which we attribute to intrusion events of bacteria into the coiled chain. 
For the polymer dynamics, we find a $t^{2}$ scaling for the center of mass mean square displacement
for intermediate times, which is absent for athermal solvent. 

We hope that our findings will stimulate new explorations. First of all, a detailed theory would be challenging which predicts at least
the scaling of the plateau behaviour. Moreover, more simulation will be necessary to understand the three-dimensional case
both for one-dimensional chains where bacteria can more easily circumvent
the polymer and for flexible membranes. In two dimensions our predictions can in principle be verified. Our model is best realized
for colloidal polymers which are built by using lock-and-key colloids as monomeric entities
\cite{Pine_Nature_2010}. These can easily be confined between glass plates and exposed to further artificial colloidal microswimmers
such that the non-crossing situations which is crucial for our set-up is realized. It is less evident how our model
is realized for real polymer chains and real bacteria as those are typical crossing in strong slit confinement. But a realted set-up
are bacteria close to a liquid-air  interface which are standardly considered in experiments, see e.g.
\cite{Marschall,Ariel,Berenike_Maier}. The latter interface is flexible but under tension.
The intrusion effect, however, is also expected to play a leading role in case the line tension is small compared to thermal effects.
Finally we think that the complex interaction between bacteria and flexible filaments as revealed in our study
may be exploited in general for the fabrication of machines on the micro- and nanoscale \cite{Trapping_OGSchmidt}.

\acknowledgments
We thank Berenike Maier, Roland Netz and Christian von Ferber for helpful discussions.
This work was financially supported by the ERC Advanced Grant INTERCOCOS
(Grant No. 267499) and by the SPP 1726 of the DFG.

\bibliography{refs}

\begin{thebibliography}{75}
\expandafter\ifx\csname natexlab\endcsname\relax\def\natexlab#1{#1}\fi
\expandafter\ifx\csname bibnamefont\endcsname\relax
  \def\bibnamefont#1{#1}\fi
\expandafter\ifx\csname bibfnamefont\endcsname\relax
  \def\bibfnamefont#1{#1}\fi
\expandafter\ifx\csname citenamefont\endcsname\relax
  \def\citenamefont#1{#1}\fi
\expandafter\ifx\csname url\endcsname\relax
  \def\url#1{\texttt{#1}}\fi
\expandafter\ifx\csname urlprefix\endcsname\relax\def\urlprefix{URL }\fi
\providecommand{\bibinfo}[2]{#2}
\providecommand{\eprint}[2][]{\url{#2}}

\bibitem[{\citenamefont{de~Gennes}(1979)}]{deGennesbook}
\bibinfo{author}{\bibfnamefont{P.~G.} \bibnamefont{de~Gennes}},
  \emph{\bibinfo{title}{Scaling concepts in polymer physics}}
  (\bibinfo{publisher}{Cornell University Press}, \bibinfo{address}{Ithaca},
  \bibinfo{year}{1979}).

\bibitem[{\citenamefont{Doi and Edwards}(1986)}]{Doi_Edwards_book}
\bibinfo{author}{\bibfnamefont{M.}~\bibnamefont{Doi}} \bibnamefont{and}
  \bibinfo{author}{\bibfnamefont{S.}~\bibnamefont{Edwards}},
  \emph{\bibinfo{title}{The Theory of Polymer Dynamics}}
  (\bibinfo{publisher}{Clarendon Press}, \bibinfo{address}{Oxford},
  \bibinfo{year}{1986}).

\bibitem[{\citenamefont{Sch\"{a}fer}(1999)}]{book_by_Lothar_Schaefer}
\bibinfo{author}{\bibfnamefont{L.}~\bibnamefont{Sch\"{a}fer}},
  \emph{\bibinfo{title}{Excluded Volume Effects in Polymer Solutions: As
  Explained by the Renormalization Group}} (\bibinfo{publisher}{Springer},
  \bibinfo{year}{1999}).

\bibitem[{\citenamefont{Sokolov et~al.}(2007)\citenamefont{Sokolov, Aranson,
  Kessler, and Goldstein}}]{2007SoEtAl}
\bibinfo{author}{\bibfnamefont{A.}~\bibnamefont{Sokolov}},
  \bibinfo{author}{\bibfnamefont{I.~S.} \bibnamefont{Aranson}},
  \bibinfo{author}{\bibfnamefont{J.~O.} \bibnamefont{Kessler}},
  \bibnamefont{and} \bibinfo{author}{\bibfnamefont{R.~E.}
  \bibnamefont{Goldstein}}, \bibinfo{journal}{Phys. Rev. Lett.}
  \textbf{\bibinfo{volume}{98}}, \bibinfo{pages}{158102}
  (\bibinfo{year}{2007}).

\bibitem[{\citenamefont{Riedel et~al.}(2005)\citenamefont{Riedel, Kruse, and
  Howard}}]{2005Riedel_Science}
\bibinfo{author}{\bibfnamefont{I.~H.} \bibnamefont{Riedel}},
  \bibinfo{author}{\bibfnamefont{K.}~\bibnamefont{Kruse}}, \bibnamefont{and}
  \bibinfo{author}{\bibfnamefont{J.}~\bibnamefont{Howard}},
  \bibinfo{journal}{Science} \textbf{\bibinfo{volume}{309}},
  \bibinfo{pages}{300} (\bibinfo{year}{2005}).

\bibitem[{\citenamefont{Vicsek and Zafeiris}(2012)}]{Vicsek_Report2012}
\bibinfo{author}{\bibfnamefont{T.}~\bibnamefont{Vicsek}} \bibnamefont{and}
  \bibinfo{author}{\bibfnamefont{A.}~\bibnamefont{Zafeiris}},
  \bibinfo{journal}{Phys. Rep.} \textbf{\bibinfo{volume}{517}},
  \bibinfo{pages}{71} (\bibinfo{year}{2012}).

\bibitem[{\citenamefont{Zhang et~al.}(2012)\citenamefont{Zhang, Klingsch,
  Schadschneider, and Seyfried}}]{Schadschneider}
\bibinfo{author}{\bibfnamefont{J.}~\bibnamefont{Zhang}},
  \bibinfo{author}{\bibfnamefont{W.}~\bibnamefont{Klingsch}},
  \bibinfo{author}{\bibfnamefont{A.}~\bibnamefont{Schadschneider}},
  \bibnamefont{and} \bibinfo{author}{\bibfnamefont{A.}~\bibnamefont{Seyfried}},
  \bibinfo{journal}{J. Stat. Mech.} \textbf{\bibinfo{volume}{2012}},
  \bibinfo{pages}{P02002} (\bibinfo{year}{2012}).

\bibitem[{\citenamefont{Paxton et~al.}(2005)\citenamefont{Paxton, Sen, and
  Mallouk}}]{Paxton}
\bibinfo{author}{\bibfnamefont{W.~F.} \bibnamefont{Paxton}},
  \bibinfo{author}{\bibfnamefont{A.}~\bibnamefont{Sen}}, \bibnamefont{and}
  \bibinfo{author}{\bibfnamefont{T.}~\bibnamefont{Mallouk}},
  \bibinfo{journal}{Chem. Eur. J.} \textbf{\bibinfo{volume}{11}},
  \bibinfo{pages}{6462} (\bibinfo{year}{2005}).

\bibitem[{\citenamefont{Snezhko and Aranson}(2011)}]{snezhko_nature}
\bibinfo{author}{\bibfnamefont{A.}~\bibnamefont{Snezhko}} \bibnamefont{and}
  \bibinfo{author}{\bibfnamefont{I.~S.} \bibnamefont{Aranson}},
  \bibinfo{journal}{Nature Mat.} \textbf{\bibinfo{volume}{10}},
  \bibinfo{pages}{698} (\bibinfo{year}{2011}).

\bibitem[{\citenamefont{K\"ummel et~al.}(2013)\citenamefont{K\"ummel, ten
  Hagen, Wittkowski, Buttinoni, Eichhorn, Volpe, L\"owen, and
  Bechinger}}]{BtH_L_part}
\bibinfo{author}{\bibfnamefont{F.}~\bibnamefont{K\"ummel}},
  \bibinfo{author}{\bibfnamefont{B.}~\bibnamefont{ten Hagen}},
  \bibinfo{author}{\bibfnamefont{R.}~\bibnamefont{Wittkowski}},
  \bibinfo{author}{\bibfnamefont{I.}~\bibnamefont{Buttinoni}},
  \bibinfo{author}{\bibfnamefont{R.}~\bibnamefont{Eichhorn}},
  \bibinfo{author}{\bibfnamefont{G.}~\bibnamefont{Volpe}},
  \bibinfo{author}{\bibfnamefont{H.}~\bibnamefont{L\"owen}}, \bibnamefont{and}
  \bibinfo{author}{\bibfnamefont{C.}~\bibnamefont{Bechinger}},
  \bibinfo{journal}{Phys. Rev. Lett.} \textbf{\bibinfo{volume}{110}},
  \bibinfo{pages}{198302} (\bibinfo{year}{2013}).

\bibitem[{\citenamefont{Palacci et~al.}(2010)\citenamefont{Palacci,
  Cottin-Bizonne, Ybert, and Bocquet}}]{BocquetPRL2010}
\bibinfo{author}{\bibfnamefont{J.}~\bibnamefont{Palacci}},
  \bibinfo{author}{\bibfnamefont{C.}~\bibnamefont{Cottin-Bizonne}},
  \bibinfo{author}{\bibfnamefont{C.}~\bibnamefont{Ybert}}, \bibnamefont{and}
  \bibinfo{author}{\bibfnamefont{L.}~\bibnamefont{Bocquet}},
  \bibinfo{journal}{Phys. Rev. Lett.} \textbf{\bibinfo{volume}{105}},
  \bibinfo{pages}{088304} (\bibinfo{year}{2010}).

\bibitem[{\citenamefont{Dreyfus et~al.}(2005)\citenamefont{Dreyfus, Baudry,
  Roper, Fermigier, Stone, and Bibette}}]{Bibette}
\bibinfo{author}{\bibfnamefont{R.}~\bibnamefont{Dreyfus}},
  \bibinfo{author}{\bibfnamefont{J.}~\bibnamefont{Baudry}},
  \bibinfo{author}{\bibfnamefont{M.~L.} \bibnamefont{Roper}},
  \bibinfo{author}{\bibfnamefont{M.}~\bibnamefont{Fermigier}},
  \bibinfo{author}{\bibfnamefont{H.~A.} \bibnamefont{Stone}}, \bibnamefont{and}
  \bibinfo{author}{\bibfnamefont{J.}~\bibnamefont{Bibette}},
  \bibinfo{journal}{Nature} \textbf{\bibinfo{volume}{437}},
  \bibinfo{pages}{862} (\bibinfo{year}{2005}).

\bibitem[{\citenamefont{Khalil et~al.}(2014)\citenamefont{Khalil, Dijkslag,
  Abelmann, and Misra}}]{MagnetoSperm}
\bibinfo{author}{\bibfnamefont{I.~S.~M.} \bibnamefont{Khalil}},
  \bibinfo{author}{\bibfnamefont{H.~C.} \bibnamefont{Dijkslag}},
  \bibinfo{author}{\bibfnamefont{L.}~\bibnamefont{Abelmann}}, \bibnamefont{and}
  \bibinfo{author}{\bibfnamefont{S.}~\bibnamefont{Misra}},
  \bibinfo{journal}{Appl. Phys. Lett.} \textbf{\bibinfo{volume}{104}},
  \bibinfo{eid}{223701} (\bibinfo{year}{2014}).

\bibitem[{\citenamefont{Popescu et~al.}(2009)\citenamefont{Popescu, Dietrich,
  and Oshanin}}]{Dietrich}
\bibinfo{author}{\bibfnamefont{M.~N.} \bibnamefont{Popescu}},
  \bibinfo{author}{\bibfnamefont{S.}~\bibnamefont{Dietrich}}, \bibnamefont{and}
  \bibinfo{author}{\bibfnamefont{G.}~\bibnamefont{Oshanin}},
  \bibinfo{journal}{J. Chem. Phys.} \textbf{\bibinfo{volume}{130}},
  \bibinfo{pages}{194702} (\bibinfo{year}{2009}).

\bibitem[{\citenamefont{S.Thakur and Kapral}(2010)}]{Kapral2010}
\bibinfo{author}{\bibnamefont{S.Thakur}} \bibnamefont{and}
  \bibinfo{author}{\bibfnamefont{R.}~\bibnamefont{Kapral}},
  \bibinfo{journal}{J. Chem. Phys.} \textbf{\bibinfo{volume}{133}},
  \bibinfo{pages}{204505} (\bibinfo{year}{2010}).

\bibitem[{\citenamefont{Volpe et~al.}(2011)\citenamefont{Volpe, Buttinoni,
  Vogt, K\"ummerer, and Bechinger}}]{Bechinger_SM11}
\bibinfo{author}{\bibfnamefont{G.}~\bibnamefont{Volpe}},
  \bibinfo{author}{\bibfnamefont{I.}~\bibnamefont{Buttinoni}},
  \bibinfo{author}{\bibfnamefont{D.}~\bibnamefont{Vogt}},
  \bibinfo{author}{\bibfnamefont{H.-J.} \bibnamefont{K\"ummerer}},
  \bibnamefont{and}
  \bibinfo{author}{\bibfnamefont{C.}~\bibnamefont{Bechinger}},
  \bibinfo{journal}{Soft Matter} \textbf{\bibinfo{volume}{7}},
  \bibinfo{pages}{8810} (\bibinfo{year}{2011}).

\bibitem[{\citenamefont{Palacci
  et~al.}(2013{\natexlab{a}})\citenamefont{Palacci, Sacanna, Vatchinsky,
  Chaikin, and Pine}}]{PalacciJACS}
\bibinfo{author}{\bibfnamefont{J.}~\bibnamefont{Palacci}},
  \bibinfo{author}{\bibfnamefont{S.}~\bibnamefont{Sacanna}},
  \bibinfo{author}{\bibfnamefont{A.}~\bibnamefont{Vatchinsky}},
  \bibinfo{author}{\bibfnamefont{P.~M.} \bibnamefont{Chaikin}},
  \bibnamefont{and} \bibinfo{author}{\bibfnamefont{D.~J.} \bibnamefont{Pine}},
  \bibinfo{journal}{J. Am. Chem. Soc.} \textbf{\bibinfo{volume}{135}},
  \bibinfo{pages}{15978} (\bibinfo{year}{2013}{\natexlab{a}}).

\bibitem[{\citenamefont{Narayan et~al.}(2007)\citenamefont{Narayan, Ramaswamy,
  and Menon}}]{Narayan}
\bibinfo{author}{\bibfnamefont{V.}~\bibnamefont{Narayan}},
  \bibinfo{author}{\bibfnamefont{S.}~\bibnamefont{Ramaswamy}},
  \bibnamefont{and} \bibinfo{author}{\bibfnamefont{N.}~\bibnamefont{Menon}},
  \bibinfo{journal}{Science} \textbf{\bibinfo{volume}{317}},
  \bibinfo{pages}{105} (\bibinfo{year}{2007}).

\bibitem[{\citenamefont{Chen et~al.}(2012)\citenamefont{Chen, Dong, Be'er,
  Swinney, and Zhang}}]{2012SwinEtAl}
\bibinfo{author}{\bibfnamefont{X.}~\bibnamefont{Chen}},
  \bibinfo{author}{\bibfnamefont{X.}~\bibnamefont{Dong}},
  \bibinfo{author}{\bibfnamefont{A.}~\bibnamefont{Be'er}},
  \bibinfo{author}{\bibfnamefont{H.~L.} \bibnamefont{Swinney}},
  \bibnamefont{and} \bibinfo{author}{\bibfnamefont{H.~P.} \bibnamefont{Zhang}},
  \bibinfo{journal}{Phys. Rev. Lett.} \textbf{\bibinfo{volume}{108}},
  \bibinfo{pages}{148101} (\bibinfo{year}{2012}).

\bibitem[{\citenamefont{Menzel and Ohta}(2012)}]{Menzel_EPL}
\bibinfo{author}{\bibfnamefont{A.~M.} \bibnamefont{Menzel}} \bibnamefont{and}
  \bibinfo{author}{\bibfnamefont{T.}~\bibnamefont{Ohta}},
  \bibinfo{journal}{Europhys. Lett.} \textbf{\bibinfo{volume}{99}},
  \bibinfo{pages}{58001} (\bibinfo{year}{2012}).

\bibitem[{\citenamefont{Gachelin et~al.}(2014)\citenamefont{Gachelin,
  Rousselet, Lindner, and Clement}}]{Clement_NJP14}
\bibinfo{author}{\bibfnamefont{J.}~\bibnamefont{Gachelin}},
  \bibinfo{author}{\bibfnamefont{A.}~\bibnamefont{Rousselet}},
  \bibinfo{author}{\bibfnamefont{A.}~\bibnamefont{Lindner}}, \bibnamefont{and}
  \bibinfo{author}{\bibfnamefont{E.}~\bibnamefont{Clement}},
  \bibinfo{journal}{New J. Phys.} \textbf{\bibinfo{volume}{16}},
  \bibinfo{pages}{025003} (\bibinfo{year}{2014}).

\bibitem[{\citenamefont{Thutupalli et~al.}(2011)\citenamefont{Thutupalli,
  Seemann, and Herminghaus}}]{2011Herminghaus}
\bibinfo{author}{\bibfnamefont{S.}~\bibnamefont{Thutupalli}},
  \bibinfo{author}{\bibfnamefont{R.}~\bibnamefont{Seemann}}, \bibnamefont{and}
  \bibinfo{author}{\bibfnamefont{S.}~\bibnamefont{Herminghaus}},
  \bibinfo{journal}{New J. Phys.} \textbf{\bibinfo{volume}{13}},
  \bibinfo{pages}{073021} (\bibinfo{year}{2011}).

\bibitem[{\citenamefont{Cohen and Golestanian}(2014)}]{Golestanian14CometSwarm}
\bibinfo{author}{\bibfnamefont{J.~A.} \bibnamefont{Cohen}} \bibnamefont{and}
  \bibinfo{author}{\bibfnamefont{R.}~\bibnamefont{Golestanian}},
  \bibinfo{journal}{Phys. Rev. Lett.} \textbf{\bibinfo{volume}{112}},
  \bibinfo{pages}{068302} (\bibinfo{year}{2014}).

\bibitem[{\citenamefont{Wensink et~al.}(2012)\citenamefont{Wensink, Dunkel,
  Heidenreich, Drescher, Goldstein, L\"owen, and Yeomans}}]{PNAS}
\bibinfo{author}{\bibfnamefont{H.~H.} \bibnamefont{Wensink}},
  \bibinfo{author}{\bibfnamefont{J.}~\bibnamefont{Dunkel}},
  \bibinfo{author}{\bibfnamefont{S.}~\bibnamefont{Heidenreich}},
  \bibinfo{author}{\bibfnamefont{K.}~\bibnamefont{Drescher}},
  \bibinfo{author}{\bibfnamefont{R.~E.} \bibnamefont{Goldstein}},
  \bibinfo{author}{\bibfnamefont{H.}~\bibnamefont{L\"owen}}, \bibnamefont{and}
  \bibinfo{author}{\bibfnamefont{J.~M.} \bibnamefont{Yeomans}},
  \bibinfo{journal}{Proc. Natl. Acad. Sci. USA} \textbf{\bibinfo{volume}{109}},
  \bibinfo{pages}{14308} (\bibinfo{year}{2012}).

\bibitem[{\citenamefont{Saintillan and Shelley}(2008)}]{2008Saint_Shelley}
\bibinfo{author}{\bibfnamefont{D.}~\bibnamefont{Saintillan}} \bibnamefont{and}
  \bibinfo{author}{\bibfnamefont{M.~J.} \bibnamefont{Shelley}},
  \bibinfo{journal}{Phys. Fluids} \textbf{\bibinfo{volume}{20}},
  \bibinfo{pages}{123304} (\bibinfo{year}{2008}).

\bibitem[{\citenamefont{Zhou et~al.}(2014)\citenamefont{Zhou, Sokolov,
  Lavrentovich, and Aranson}}]{ISA-PNAS14}
\bibinfo{author}{\bibfnamefont{S.}~\bibnamefont{Zhou}},
  \bibinfo{author}{\bibfnamefont{A.}~\bibnamefont{Sokolov}},
  \bibinfo{author}{\bibfnamefont{O.~D.} \bibnamefont{Lavrentovich}},
  \bibnamefont{and} \bibinfo{author}{\bibfnamefont{I.~S.}
  \bibnamefont{Aranson}}, \bibinfo{journal}{Proc. Natl. Acad. Sci. USA}
  \textbf{\bibinfo{volume}{111}}, \bibinfo{pages}{1265} (\bibinfo{year}{2014}).

\bibitem[{\citenamefont{Liu and I}(2013)}]{Li_PRE}
\bibinfo{author}{\bibfnamefont{K.-A.} \bibnamefont{Liu}} \bibnamefont{and}
  \bibinfo{author}{\bibfnamefont{L.}~\bibnamefont{I}}, \bibinfo{journal}{Phys.
  Rev. E} \textbf{\bibinfo{volume}{88}}, \bibinfo{pages}{033004}
  (\bibinfo{year}{2013}).

\bibitem[{\citenamefont{Palacci
  et~al.}(2013{\natexlab{b}})\citenamefont{Palacci, Sacanna, Steinberg, Pine,
  and Chaikin}}]{Palacci_science}
\bibinfo{author}{\bibfnamefont{J.}~\bibnamefont{Palacci}},
  \bibinfo{author}{\bibfnamefont{S.}~\bibnamefont{Sacanna}},
  \bibinfo{author}{\bibfnamefont{A.~P.} \bibnamefont{Steinberg}},
  \bibinfo{author}{\bibfnamefont{D.~J.} \bibnamefont{Pine}}, \bibnamefont{and}
  \bibinfo{author}{\bibfnamefont{P.~M.} \bibnamefont{Chaikin}},
  \bibinfo{journal}{Science} \textbf{\bibinfo{volume}{339}},
  \bibinfo{pages}{936} (\bibinfo{year}{2013}{\natexlab{b}}).

\bibitem[{\citenamefont{Buttinoni et~al.}(2013)\citenamefont{Buttinoni,
  Bialk\'e, K\"ummel, L\"owen, Bechinger, and Speck}}]{Bialke_PRL2013}
\bibinfo{author}{\bibfnamefont{I.}~\bibnamefont{Buttinoni}},
  \bibinfo{author}{\bibfnamefont{J.}~\bibnamefont{Bialk\'e}},
  \bibinfo{author}{\bibfnamefont{F.}~\bibnamefont{K\"ummel}},
  \bibinfo{author}{\bibfnamefont{H.}~\bibnamefont{L\"owen}},
  \bibinfo{author}{\bibfnamefont{C.}~\bibnamefont{Bechinger}},
  \bibnamefont{and} \bibinfo{author}{\bibfnamefont{T.}~\bibnamefont{Speck}},
  \bibinfo{journal}{Phys. Rev. Lett.} \textbf{\bibinfo{volume}{110}},
  \bibinfo{pages}{238301} (\bibinfo{year}{2013}).

\bibitem[{\citenamefont{Redner et~al.}(2013)\citenamefont{Redner, Hagan, and
  Baskaran}}]{Baskaran_PRL2013}
\bibinfo{author}{\bibfnamefont{G.~S.} \bibnamefont{Redner}},
  \bibinfo{author}{\bibfnamefont{M.~F.} \bibnamefont{Hagan}}, \bibnamefont{and}
  \bibinfo{author}{\bibfnamefont{A.}~\bibnamefont{Baskaran}},
  \bibinfo{journal}{Phys. Rev. Lett.} \textbf{\bibinfo{volume}{110}},
  \bibinfo{pages}{055701} (\bibinfo{year}{2013}).

\bibitem[{\citenamefont{Fily and Marchetti}(2012)}]{MarchettiPRL12}
\bibinfo{author}{\bibfnamefont{Y.}~\bibnamefont{Fily}} \bibnamefont{and}
  \bibinfo{author}{\bibfnamefont{M.~C.} \bibnamefont{Marchetti}},
  \bibinfo{journal}{Phys. Rev. Lett.} \textbf{\bibinfo{volume}{108}},
  \bibinfo{pages}{235702} (\bibinfo{year}{2012}).

\bibitem[{\citenamefont{Z\"ottl and Stark}(2014)}]{ZoettelPRL14}
\bibinfo{author}{\bibfnamefont{A.}~\bibnamefont{Z\"ottl}} \bibnamefont{and}
  \bibinfo{author}{\bibfnamefont{H.}~\bibnamefont{Stark}},
  \bibinfo{journal}{Phys. Rev. Lett.} \textbf{\bibinfo{volume}{112}},
  \bibinfo{pages}{118101} (\bibinfo{year}{2014}).

\bibitem[{\citenamefont{Das et~al.}(2014)\citenamefont{Das, Egorov, Trefz,
  Virnau, and Binder}}]{Egorov}
\bibinfo{author}{\bibfnamefont{S.~K.} \bibnamefont{Das}},
  \bibinfo{author}{\bibfnamefont{S.~A.} \bibnamefont{Egorov}},
  \bibinfo{author}{\bibfnamefont{B.}~\bibnamefont{Trefz}},
  \bibinfo{author}{\bibfnamefont{P.}~\bibnamefont{Virnau}}, \bibnamefont{and}
  \bibinfo{author}{\bibfnamefont{K.}~\bibnamefont{Binder}},
  \bibinfo{journal}{Phys. Rev. Lett.} \textbf{\bibinfo{volume}{112}},
  \bibinfo{pages}{198301} (\bibinfo{year}{2014}).

\bibitem[{\citenamefont{Wensink and L\"owen}(2008)}]{Wensink2008}
\bibinfo{author}{\bibfnamefont{H.~H.} \bibnamefont{Wensink}} \bibnamefont{and}
  \bibinfo{author}{\bibfnamefont{H.}~\bibnamefont{L\"owen}},
  \bibinfo{journal}{Phys. Rev. E} \textbf{\bibinfo{volume}{78}},
  \bibinfo{pages}{031409} (\bibinfo{year}{2008}).

\bibitem[{\citenamefont{Wan et~al.}(2008)\citenamefont{Wan, Olson~Reichhardt,
  Nussinov, and Reichhardt}}]{Reichhardt_PRL2008}
\bibinfo{author}{\bibfnamefont{M.~B.} \bibnamefont{Wan}},
  \bibinfo{author}{\bibfnamefont{C.~J.} \bibnamefont{Olson~Reichhardt}},
  \bibinfo{author}{\bibfnamefont{Z.}~\bibnamefont{Nussinov}}, \bibnamefont{and}
  \bibinfo{author}{\bibfnamefont{C.}~\bibnamefont{Reichhardt}},
  \bibinfo{journal}{Phys. Rev. Lett.} \textbf{\bibinfo{volume}{101}},
  \bibinfo{pages}{018102} (\bibinfo{year}{2008}).

\bibitem[{\citenamefont{Berdakin et~al.}(2013)\citenamefont{Berdakin, Jeyaram,
  Moshchalkov, Venken, Dierckx, Vanderleyden, Silhanek, Condat, and
  Marconi}}]{MarconiPRE}
\bibinfo{author}{\bibfnamefont{I.}~\bibnamefont{Berdakin}},
  \bibinfo{author}{\bibfnamefont{Y.}~\bibnamefont{Jeyaram}},
  \bibinfo{author}{\bibfnamefont{V.~V.} \bibnamefont{Moshchalkov}},
  \bibinfo{author}{\bibfnamefont{L.}~\bibnamefont{Venken}},
  \bibinfo{author}{\bibfnamefont{S.}~\bibnamefont{Dierckx}},
  \bibinfo{author}{\bibfnamefont{S.~J.} \bibnamefont{Vanderleyden}},
  \bibinfo{author}{\bibfnamefont{A.~V.} \bibnamefont{Silhanek}},
  \bibinfo{author}{\bibfnamefont{C.~A.} \bibnamefont{Condat}},
  \bibnamefont{and} \bibinfo{author}{\bibfnamefont{V.~I.}
  \bibnamefont{Marconi}}, \bibinfo{journal}{Phys. Rev. E}
  \textbf{\bibinfo{volume}{87}}, \bibinfo{pages}{052702}
  (\bibinfo{year}{2013}).

\bibitem[{\citenamefont{Elgeti and Gompper}(2013)}]{ElgetiGompper13}
\bibinfo{author}{\bibfnamefont{J.}~\bibnamefont{Elgeti}} \bibnamefont{and}
  \bibinfo{author}{\bibfnamefont{G.}~\bibnamefont{Gompper}},
  \bibinfo{journal}{Europhys. Lett.} \textbf{\bibinfo{volume}{101}},
  \bibinfo{pages}{48003} (\bibinfo{year}{2013}).

\bibitem[{\citenamefont{Lee}(2013)}]{Lee13Wall}
\bibinfo{author}{\bibfnamefont{C.~F.} \bibnamefont{Lee}}, \bibinfo{journal}{New
  J. Phys.} \textbf{\bibinfo{volume}{15}}, \bibinfo{pages}{055007}
  (\bibinfo{year}{2013}).

\bibitem[{\citenamefont{Fily et~al.}(2014)\citenamefont{Fily, Baskaran, and
  Hagan}}]{HaganConfinement}
\bibinfo{author}{\bibfnamefont{Y.}~\bibnamefont{Fily}},
  \bibinfo{author}{\bibfnamefont{A.}~\bibnamefont{Baskaran}}, \bibnamefont{and}
  \bibinfo{author}{\bibfnamefont{M.~F.} \bibnamefont{Hagan}},
  \bibinfo{journal}{arXiv preprint, arXiv:1402.5583}  (\bibinfo{year}{2014}).

\bibitem[{\citenamefont{Galajda et~al.}(2007)\citenamefont{Galajda, Keymer,
  Chaikin, and Austin}}]{Chaikin2007}
\bibinfo{author}{\bibfnamefont{P.}~\bibnamefont{Galajda}},
  \bibinfo{author}{\bibfnamefont{J.}~\bibnamefont{Keymer}},
  \bibinfo{author}{\bibfnamefont{P.}~\bibnamefont{Chaikin}}, \bibnamefont{and}
  \bibinfo{author}{\bibfnamefont{R.}~\bibnamefont{Austin}},
  \bibinfo{journal}{J. Bacteriol.} \textbf{\bibinfo{volume}{189}},
  \bibinfo{pages}{8704} (\bibinfo{year}{2007}).

\bibitem[{\citenamefont{Guidobaldi et~al.}(2014)\citenamefont{Guidobaldi,
  Jeyaram, Berdakin, Moshchalkov, Condat, Marconi, Giojalas, and
  Silhanek}}]{TrappingSperms}
\bibinfo{author}{\bibfnamefont{A.}~\bibnamefont{Guidobaldi}},
  \bibinfo{author}{\bibfnamefont{Y.}~\bibnamefont{Jeyaram}},
  \bibinfo{author}{\bibfnamefont{I.}~\bibnamefont{Berdakin}},
  \bibinfo{author}{\bibfnamefont{V.~V.} \bibnamefont{Moshchalkov}},
  \bibinfo{author}{\bibfnamefont{C.~A.} \bibnamefont{Condat}},
  \bibinfo{author}{\bibfnamefont{V.~I.} \bibnamefont{Marconi}},
  \bibinfo{author}{\bibfnamefont{L.}~\bibnamefont{Giojalas}}, \bibnamefont{and}
  \bibinfo{author}{\bibfnamefont{A.~V.} \bibnamefont{Silhanek}},
  \bibinfo{journal}{Phys. Rev. E} \textbf{\bibinfo{volume}{89}},
  \bibinfo{pages}{032720} (\bibinfo{year}{2014}).

\bibitem[{\citenamefont{Kaiser et~al.}(2012)\citenamefont{Kaiser, Wensink, and
  L\"owen}}]{Kaiser_PRL}
\bibinfo{author}{\bibfnamefont{A.}~\bibnamefont{Kaiser}},
  \bibinfo{author}{\bibfnamefont{H.~H.} \bibnamefont{Wensink}},
  \bibnamefont{and} \bibinfo{author}{\bibfnamefont{H.}~\bibnamefont{L\"owen}},
  \bibinfo{journal}{Phys. Rev. Lett.} \textbf{\bibinfo{volume}{108}},
  \bibinfo{pages}{268307} (\bibinfo{year}{2012}).

\bibitem[{\citenamefont{Tailleur and Cates}(2009)}]{CatesTrapping}
\bibinfo{author}{\bibfnamefont{J.}~\bibnamefont{Tailleur}} \bibnamefont{and}
  \bibinfo{author}{\bibfnamefont{M.~E.} \bibnamefont{Cates}},
  \bibinfo{journal}{Europhys. Lett.} \textbf{\bibinfo{volume}{86}},
  \bibinfo{pages}{60002} (\bibinfo{year}{2009}).

\bibitem[{\citenamefont{Chepizhko and Peruani}(2013)}]{TrappingPeruani}
\bibinfo{author}{\bibfnamefont{O.}~\bibnamefont{Chepizhko}} \bibnamefont{and}
  \bibinfo{author}{\bibfnamefont{F.}~\bibnamefont{Peruani}},
  \bibinfo{journal}{Phys. Rev. Lett.} \textbf{\bibinfo{volume}{111}},
  \bibinfo{pages}{160604} (\bibinfo{year}{2013}).

\bibitem[{\citenamefont{Restrepo-Perez
  et~al.}(2014)\citenamefont{Restrepo-Perez, Soler, Martinez-Cisneros, Sanchez,
  and Schmidt}}]{Trapping_OGSchmidt}
\bibinfo{author}{\bibfnamefont{L.}~\bibnamefont{Restrepo-Perez}},
  \bibinfo{author}{\bibfnamefont{L.}~\bibnamefont{Soler}},
  \bibinfo{author}{\bibfnamefont{C.~S.} \bibnamefont{Martinez-Cisneros}},
  \bibinfo{author}{\bibfnamefont{S.}~\bibnamefont{Sanchez}}, \bibnamefont{and}
  \bibinfo{author}{\bibfnamefont{O.~G.} \bibnamefont{Schmidt}},
  \bibinfo{journal}{Lab Chip} \textbf{\bibinfo{volume}{14}},
  \bibinfo{pages}{1515} (\bibinfo{year}{2014}).

\bibitem[{\citenamefont{Kaiser et~al.}(2014)\citenamefont{Kaiser, Peshkov,
  Sokolov, ten Hagen, L\"owen, and Aranson}}]{KaiserSokolov_2014}
\bibinfo{author}{\bibfnamefont{A.}~\bibnamefont{Kaiser}},
  \bibinfo{author}{\bibfnamefont{A.}~\bibnamefont{Peshkov}},
  \bibinfo{author}{\bibfnamefont{A.}~\bibnamefont{Sokolov}},
  \bibinfo{author}{\bibfnamefont{B.}~\bibnamefont{ten Hagen}},
  \bibinfo{author}{\bibfnamefont{H.}~\bibnamefont{L\"owen}}, \bibnamefont{and}
  \bibinfo{author}{\bibfnamefont{I.~S.} \bibnamefont{Aranson}},
  \bibinfo{journal}{Phys. Rev. Lett.} \textbf{\bibinfo{volume}{112}},
  \bibinfo{pages}{158101} (\bibinfo{year}{2014}).

\bibitem[{\citenamefont{Sokolov et~al.}(2010)\citenamefont{Sokolov, Apodaca,
  Grzyboski, and Aranson}}]{SokolovPNAS}
\bibinfo{author}{\bibfnamefont{A.}~\bibnamefont{Sokolov}},
  \bibinfo{author}{\bibfnamefont{M.~M.} \bibnamefont{Apodaca}},
  \bibinfo{author}{\bibfnamefont{B.~A.} \bibnamefont{Grzyboski}},
  \bibnamefont{and} \bibinfo{author}{\bibfnamefont{I.~S.}
  \bibnamefont{Aranson}}, \bibinfo{journal}{Proc. Natl. Acad. Sci. USA}
  \textbf{\bibinfo{volume}{107}}, \bibinfo{pages}{969} (\bibinfo{year}{2010}).

\bibitem[{\citenamefont{DiLeonardo et~al.}(2010)\citenamefont{DiLeonardo,
  Angelani, DellArciprete, Ruocco, Iebba, Schippa, Conte, Mecarini, Angelis,
  and Fabrizio}}]{LeonardoPNAS}
\bibinfo{author}{\bibfnamefont{R.}~\bibnamefont{DiLeonardo}},
  \bibinfo{author}{\bibfnamefont{L.}~\bibnamefont{Angelani}},
  \bibinfo{author}{\bibfnamefont{D.}~\bibnamefont{DellArciprete}},
  \bibinfo{author}{\bibfnamefont{G.}~\bibnamefont{Ruocco}},
  \bibinfo{author}{\bibfnamefont{V.}~\bibnamefont{Iebba}},
  \bibinfo{author}{\bibfnamefont{S.}~\bibnamefont{Schippa}},
  \bibinfo{author}{\bibfnamefont{M.~P.} \bibnamefont{Conte}},
  \bibinfo{author}{\bibfnamefont{F.}~\bibnamefont{Mecarini}},
  \bibinfo{author}{\bibfnamefont{F.~D.} \bibnamefont{Angelis}},
  \bibnamefont{and} \bibinfo{author}{\bibfnamefont{E.~D.}
  \bibnamefont{Fabrizio}}, \bibinfo{journal}{Proc. Natl. Acad. Sci. USA}
  \textbf{\bibinfo{volume}{107}}, \bibinfo{pages}{9541} (\bibinfo{year}{2010}).

\bibitem[{\citenamefont{Schulz et~al.}(2012)\citenamefont{Schulz, Schmidt,
  Best, Dzubiella, and Netz}}]{Netz2012}
\bibinfo{author}{\bibfnamefont{J.~C.~F.} \bibnamefont{Schulz}},
  \bibinfo{author}{\bibfnamefont{L.}~\bibnamefont{Schmidt}},
  \bibinfo{author}{\bibfnamefont{R.~B.} \bibnamefont{Best}},
  \bibinfo{author}{\bibfnamefont{J.}~\bibnamefont{Dzubiella}},
  \bibnamefont{and} \bibinfo{author}{\bibfnamefont{R.~R.} \bibnamefont{Netz}},
  \bibinfo{journal}{J. Am. Chem. Soc.} \textbf{\bibinfo{volume}{134}},
  \bibinfo{pages}{6273} (\bibinfo{year}{2012}).

\bibitem[{\citenamefont{Winkler}(2006)}]{WnklerPolyShearFlow06}
\bibinfo{author}{\bibfnamefont{R.~G.} \bibnamefont{Winkler}},
  \bibinfo{journal}{Phys. Rev. Lett.} \textbf{\bibinfo{volume}{97}},
  \bibinfo{pages}{128301} (\bibinfo{year}{2006}).

\bibitem[{\citenamefont{Liu et~al.}(2004)\citenamefont{Liu, Ashok, and
  Muthukumar}}]{Muthukumar2004}
\bibinfo{author}{\bibfnamefont{S.}~\bibnamefont{Liu}},
  \bibinfo{author}{\bibfnamefont{B.}~\bibnamefont{Ashok}}, \bibnamefont{and}
  \bibinfo{author}{\bibfnamefont{M.}~\bibnamefont{Muthukumar}},
  \bibinfo{journal}{Polymer} \textbf{\bibinfo{volume}{45}},
  \bibinfo{pages}{1383 } (\bibinfo{year}{2004}).

\bibitem[{\citenamefont{Cannavacciuolo
  et~al.}(2008)\citenamefont{Cannavacciuolo, Winkler, and
  Gompper}}]{WinklerGompper2008}
\bibinfo{author}{\bibfnamefont{L.}~\bibnamefont{Cannavacciuolo}},
  \bibinfo{author}{\bibfnamefont{R.~G.} \bibnamefont{Winkler}},
  \bibnamefont{and} \bibinfo{author}{\bibfnamefont{G.}~\bibnamefont{Gompper}},
  \bibinfo{journal}{Europhys. Lett.} \textbf{\bibinfo{volume}{83}},
  \bibinfo{pages}{34007} (\bibinfo{year}{2008}).

\bibitem[{\citenamefont{Kaiser et~al.}(2013)\citenamefont{Kaiser, Popowa,
  Wensink, and L\"owen}}]{KaiserPopowa_2013}
\bibinfo{author}{\bibfnamefont{A.}~\bibnamefont{Kaiser}},
  \bibinfo{author}{\bibfnamefont{K.}~\bibnamefont{Popowa}},
  \bibinfo{author}{\bibfnamefont{H.~H.} \bibnamefont{Wensink}},
  \bibnamefont{and} \bibinfo{author}{\bibfnamefont{H.}~\bibnamefont{L\"owen}},
  \bibinfo{journal}{Phys. Rev. E} \textbf{\bibinfo{volume}{88}},
  \bibinfo{pages}{022311} (\bibinfo{year}{2013}).

\bibitem[{\citenamefont{Angelani and Leonardo}(2010)}]{AngelaniCARGO}
\bibinfo{author}{\bibfnamefont{L.}~\bibnamefont{Angelani}} \bibnamefont{and}
  \bibinfo{author}{\bibfnamefont{R.~D.} \bibnamefont{Leonardo}},
  \bibinfo{journal}{New J. Phys.} \textbf{\bibinfo{volume}{12}},
  \bibinfo{pages}{113017} (\bibinfo{year}{2010}).

\bibitem[{\citenamefont{Ledesma-Aguilar and Yeomans}(2013)}]{ElasticConf}
\bibinfo{author}{\bibfnamefont{R.}~\bibnamefont{Ledesma-Aguilar}}
  \bibnamefont{and} \bibinfo{author}{\bibfnamefont{J.~M.}
  \bibnamefont{Yeomans}}, \bibinfo{journal}{Phys. Rev. Lett.}
  \textbf{\bibinfo{volume}{111}}, \bibinfo{pages}{138101}
  (\bibinfo{year}{2013}).

\bibitem[{\citenamefont{Kjelleberg et~al.}(1982)\citenamefont{Kjelleberg,
  Humphrey, and Marshall}}]{Marschall}
\bibinfo{author}{\bibfnamefont{S.}~\bibnamefont{Kjelleberg}},
  \bibinfo{author}{\bibfnamefont{B.~A.} \bibnamefont{Humphrey}},
  \bibnamefont{and} \bibinfo{author}{\bibfnamefont{K.~C.}
  \bibnamefont{Marshall}}, \bibinfo{journal}{Appl. Environ. Microbiol.}
  \textbf{\bibinfo{volume}{43}}, \bibinfo{pages}{1166} (\bibinfo{year}{1982}).

\bibitem[{\citenamefont{Pitt et~al.}(1993)\citenamefont{Pitt, McBride, Barton,
  and Sagers}}]{Pitt}
\bibinfo{author}{\bibfnamefont{W.~G.} \bibnamefont{Pitt}},
  \bibinfo{author}{\bibfnamefont{M.~O.} \bibnamefont{McBride}},
  \bibinfo{author}{\bibfnamefont{A.~J.} \bibnamefont{Barton}},
  \bibnamefont{and} \bibinfo{author}{\bibfnamefont{R.~D.}
  \bibnamefont{Sagers}}, \bibinfo{journal}{Biomaterials}
  \textbf{\bibinfo{volume}{14}}, \bibinfo{pages}{605 } (\bibinfo{year}{1993}).

\bibitem[{\citenamefont{G\'omez-Su\'arez
  et~al.}(2001)\citenamefont{G\'omez-Su\'arez, Busscher, and van~der
  Mei}}]{VanDerMei}
\bibinfo{author}{\bibfnamefont{C.}~\bibnamefont{G\'omez-Su\'arez}},
  \bibinfo{author}{\bibfnamefont{H.~J.} \bibnamefont{Busscher}},
  \bibnamefont{and} \bibinfo{author}{\bibfnamefont{H.~C.} \bibnamefont{van~der
  Mei}}, \bibinfo{journal}{Appl. Environ. Microbiol.}
  \textbf{\bibinfo{volume}{67}}, \bibinfo{pages}{2531} (\bibinfo{year}{2001}).

\bibitem[{\citenamefont{Powelson and Mills}(1996)}]{Mills}
\bibinfo{author}{\bibfnamefont{D.~K.} \bibnamefont{Powelson}} \bibnamefont{and}
  \bibinfo{author}{\bibfnamefont{A.~L.} \bibnamefont{Mills}},
  \bibinfo{journal}{Appl. Environ. Microbiol.} \textbf{\bibinfo{volume}{62}},
  \bibinfo{pages}{2593} (\bibinfo{year}{1996}).

\bibitem[{\citenamefont{Meel et~al.}(2012)\citenamefont{Meel, Kouzel,
  Oldewurtel, and Maier}}]{Berenike_Maier}
\bibinfo{author}{\bibfnamefont{C.}~\bibnamefont{Meel}},
  \bibinfo{author}{\bibfnamefont{N.}~\bibnamefont{Kouzel}},
  \bibinfo{author}{\bibfnamefont{E.~R.} \bibnamefont{Oldewurtel}},
  \bibnamefont{and} \bibinfo{author}{\bibfnamefont{B.}~\bibnamefont{Maier}},
  \bibinfo{journal}{Small} \textbf{\bibinfo{volume}{8}}, \bibinfo{pages}{530}
  (\bibinfo{year}{2012}).

\bibitem[{\citenamefont{Shklarsh et~al.}(2012)\citenamefont{Shklarsh,
  Finkelshtein, Ariel, Kalisman, Ingham, and Ben-Jacob}}]{Ariel}
\bibinfo{author}{\bibfnamefont{A.}~\bibnamefont{Shklarsh}},
  \bibinfo{author}{\bibfnamefont{A.}~\bibnamefont{Finkelshtein}},
  \bibinfo{author}{\bibfnamefont{G.}~\bibnamefont{Ariel}},
  \bibinfo{author}{\bibfnamefont{O.}~\bibnamefont{Kalisman}},
  \bibinfo{author}{\bibfnamefont{C.}~\bibnamefont{Ingham}}, \bibnamefont{and}
  \bibinfo{author}{\bibfnamefont{E.}~\bibnamefont{Ben-Jacob}},
  \bibinfo{journal}{Interface Focus} \textbf{\bibinfo{volume}{2}},
  \bibinfo{pages}{689} (\bibinfo{year}{2012}).

\bibitem[{\citenamefont{Sacanna et~al.}(2010)\citenamefont{Sacanna, Irvine,
  Chaikin, and Pine}}]{Pine_Nature_2010}
\bibinfo{author}{\bibfnamefont{S.}~\bibnamefont{Sacanna}},
  \bibinfo{author}{\bibfnamefont{W.~T.~M.} \bibnamefont{Irvine}},
  \bibinfo{author}{\bibfnamefont{P.~M.} \bibnamefont{Chaikin}},
  \bibnamefont{and} \bibinfo{author}{\bibfnamefont{D.~J.} \bibnamefont{Pine}},
  \bibinfo{journal}{Nature} \textbf{\bibinfo{volume}{464}},
  \bibinfo{pages}{575} (\bibinfo{year}{2010}).

\bibitem[{\citenamefont{Wang et~al.}(2013)\citenamefont{Wang, Duan, Sen, and
  Mallouk}}]{SenPNAS13}
\bibinfo{author}{\bibfnamefont{W.}~\bibnamefont{Wang}},
  \bibinfo{author}{\bibfnamefont{W.}~\bibnamefont{Duan}},
  \bibinfo{author}{\bibfnamefont{A.}~\bibnamefont{Sen}}, \bibnamefont{and}
  \bibinfo{author}{\bibfnamefont{T.~E.} \bibnamefont{Mallouk}},
  \bibinfo{journal}{Proc. Natl. Acad. Sci. USA} \textbf{\bibinfo{volume}{110}},
  \bibinfo{pages}{17744} (\bibinfo{year}{2013}).

\bibitem[{\citenamefont{Buttinoni et~al.}(2012)\citenamefont{Buttinoni, Volpe,
  K\"ummel, Volpe, and Bechinger}}]{Bechinger}
\bibinfo{author}{\bibfnamefont{I.}~\bibnamefont{Buttinoni}},
  \bibinfo{author}{\bibfnamefont{G.}~\bibnamefont{Volpe}},
  \bibinfo{author}{\bibfnamefont{F.}~\bibnamefont{K\"ummel}},
  \bibinfo{author}{\bibfnamefont{G.}~\bibnamefont{Volpe}}, \bibnamefont{and}
  \bibinfo{author}{\bibfnamefont{C.}~\bibnamefont{Bechinger}},
  \bibinfo{journal}{J. Phys. Condens. Matter} \textbf{\bibinfo{volume}{24}},
  \bibinfo{pages}{284129} (\bibinfo{year}{2012}).

\bibitem[{\citenamefont{Wu et~al.}(1992)\citenamefont{Wu, Hui, and
  Chandler}}]{ChandlerJCP92}
\bibinfo{author}{\bibfnamefont{D.}~\bibnamefont{Wu}},
  \bibinfo{author}{\bibfnamefont{K.}~\bibnamefont{Hui}}, \bibnamefont{and}
  \bibinfo{author}{\bibfnamefont{D.}~\bibnamefont{Chandler}},
  \bibinfo{journal}{J. Chem. Phys.} \textbf{\bibinfo{volume}{96}},
  \bibinfo{pages}{835} (\bibinfo{year}{1992}).

\bibitem[{\citenamefont{Leung and Chandler}(1995)}]{ChandlerJCP95}
\bibinfo{author}{\bibfnamefont{K.}~\bibnamefont{Leung}} \bibnamefont{and}
  \bibinfo{author}{\bibfnamefont{D.}~\bibnamefont{Chandler}},
  \bibinfo{journal}{J. Chem. Phys.} \textbf{\bibinfo{volume}{102}},
  \bibinfo{pages}{1405} (\bibinfo{year}{1995}).

\bibitem[{\citenamefont{Kremer and Grest}(1990)}]{KremerMDPolymer1990}
\bibinfo{author}{\bibfnamefont{K.}~\bibnamefont{Kremer}} \bibnamefont{and}
  \bibinfo{author}{\bibfnamefont{G.~S.} \bibnamefont{Grest}},
  \bibinfo{journal}{J. Chem. Phys.} \textbf{\bibinfo{volume}{92}},
  \bibinfo{pages}{5057} (\bibinfo{year}{1990}).

\bibitem[{\citenamefont{Wensink and L\"owen}(2012)}]{WensinkJPCM}
\bibinfo{author}{\bibfnamefont{H.~H.} \bibnamefont{Wensink}} \bibnamefont{and}
  \bibinfo{author}{\bibfnamefont{H.}~\bibnamefont{L\"owen}},
  \bibinfo{journal}{J. Phys. Condens. Matter} \textbf{\bibinfo{volume}{24}},
  \bibinfo{pages}{464130} (\bibinfo{year}{2012}).

\bibitem[{\citenamefont{Bialk\'e et~al.}(2012)\citenamefont{Bialk\'e, Speck,
  and L\"owen}}]{Bialke_PRL2012}
\bibinfo{author}{\bibfnamefont{J.}~\bibnamefont{Bialk\'e}},
  \bibinfo{author}{\bibfnamefont{T.}~\bibnamefont{Speck}}, \bibnamefont{and}
  \bibinfo{author}{\bibfnamefont{H.}~\bibnamefont{L\"owen}},
  \bibinfo{journal}{Phys. Rev. Lett.} \textbf{\bibinfo{volume}{108}},
  \bibinfo{pages}{168301} (\bibinfo{year}{2012}).

\bibitem[{\citenamefont{Hinczewski et~al.}(2009)\citenamefont{Hinczewski,
  Schlagberger, Rubinstein, Krichevsky, and Netz}}]{NetzEndmonomer}
\bibinfo{author}{\bibfnamefont{M.}~\bibnamefont{Hinczewski}},
  \bibinfo{author}{\bibfnamefont{X.}~\bibnamefont{Schlagberger}},
  \bibinfo{author}{\bibfnamefont{M.}~\bibnamefont{Rubinstein}},
  \bibinfo{author}{\bibfnamefont{O.}~\bibnamefont{Krichevsky}},
  \bibnamefont{and} \bibinfo{author}{\bibfnamefont{R.~R.} \bibnamefont{Netz}},
  \bibinfo{journal}{Macromolecules} \textbf{\bibinfo{volume}{42}},
  \bibinfo{pages}{860} (\bibinfo{year}{2009}).

\bibitem[{\citenamefont{ten Hagen et~al.}(2011)\citenamefont{ten Hagen, van
  Teeffelen, and L\"owen}}]{BtH2011}
\bibinfo{author}{\bibfnamefont{B.}~\bibnamefont{ten Hagen}},
  \bibinfo{author}{\bibfnamefont{S.}~\bibnamefont{van Teeffelen}},
  \bibnamefont{and} \bibinfo{author}{\bibfnamefont{H.}~\bibnamefont{L\"owen}},
  \bibinfo{journal}{J. Phys. Condens. Matter} \textbf{\bibinfo{volume}{23}},
  \bibinfo{pages}{194119} (\bibinfo{year}{2011}).

\bibitem[{\citenamefont{Howse et~al.}(2007)\citenamefont{Howse, Jones, Ryan,
  Gough, Vafabakhsh, and Golestanian}}]{Howse_2007}
\bibinfo{author}{\bibfnamefont{J.~R.} \bibnamefont{Howse}},
  \bibinfo{author}{\bibfnamefont{R.~A.~L.} \bibnamefont{Jones}},
  \bibinfo{author}{\bibfnamefont{A.~J.} \bibnamefont{Ryan}},
  \bibinfo{author}{\bibfnamefont{T.}~\bibnamefont{Gough}},
  \bibinfo{author}{\bibfnamefont{R.}~\bibnamefont{Vafabakhsh}},
  \bibnamefont{and}
  \bibinfo{author}{\bibfnamefont{R.}~\bibnamefont{Golestanian}},
  \bibinfo{journal}{Phys. Rev. Lett.} \textbf{\bibinfo{volume}{99}},
  \bibinfo{pages}{048102} (\bibinfo{year}{2007}).

\bibitem[{\citenamefont{Zheng et~al.}(2013)\citenamefont{Zheng, ten Hagen,
  Kaiser, Wu, Cui, Silber-Li, and L\"owen}}]{Kaiser_PRE2013Janus}
\bibinfo{author}{\bibfnamefont{X.}~\bibnamefont{Zheng}},
  \bibinfo{author}{\bibfnamefont{B.}~\bibnamefont{ten Hagen}},
  \bibinfo{author}{\bibfnamefont{A.}~\bibnamefont{Kaiser}},
  \bibinfo{author}{\bibfnamefont{M.}~\bibnamefont{Wu}},
  \bibinfo{author}{\bibfnamefont{H.}~\bibnamefont{Cui}},
  \bibinfo{author}{\bibfnamefont{Z.}~\bibnamefont{Silber-Li}},
  \bibnamefont{and} \bibinfo{author}{\bibfnamefont{H.}~\bibnamefont{L\"owen}},
  \bibinfo{journal}{Phys. Rev. E} \textbf{\bibinfo{volume}{88}},
  \bibinfo{pages}{032304} (\bibinfo{year}{2013}).

\bibitem[{\citenamefont{Leptos et~al.}(2009)\citenamefont{Leptos, Guasto,
  Gollub, Pesci, and Goldstein}}]{TracerGoldstein}
\bibinfo{author}{\bibfnamefont{K.~C.} \bibnamefont{Leptos}},
  \bibinfo{author}{\bibfnamefont{J.~S.} \bibnamefont{Guasto}},
  \bibinfo{author}{\bibfnamefont{J.~P.} \bibnamefont{Gollub}},
  \bibinfo{author}{\bibfnamefont{A.~I.} \bibnamefont{Pesci}}, \bibnamefont{and}
  \bibinfo{author}{\bibfnamefont{R.~E.} \bibnamefont{Goldstein}},
  \bibinfo{journal}{Phys. Rev. Lett.} \textbf{\bibinfo{volume}{103}},
  \bibinfo{pages}{198103} (\bibinfo{year}{2009}).

\bibitem[{\citenamefont{Mi\~no et~al.}(2013)\citenamefont{Mi\~no, Dunstan,
  Rousselet, Clement, and Soto}}]{TracerClement}
\bibinfo{author}{\bibfnamefont{G.}~\bibnamefont{Mi\~no}},
  \bibinfo{author}{\bibfnamefont{J.}~\bibnamefont{Dunstan}},
  \bibinfo{author}{\bibfnamefont{A.}~\bibnamefont{Rousselet}},
  \bibinfo{author}{\bibfnamefont{E.}~\bibnamefont{Clement}}, \bibnamefont{and}
  \bibinfo{author}{\bibfnamefont{R.}~\bibnamefont{Soto}}, \bibinfo{journal}{J.
  Fluid Mech.} \textbf{\bibinfo{volume}{729}}, \bibinfo{pages}{423}
  (\bibinfo{year}{2013}).

\end{thebibliography}

\end{document}